\documentclass[aps,pra,letterpaper,twocolumn,showpacs,,amsmath,amssymb,superscriptaddress]{revtex4-1}
\usepackage{graphicx}
\usepackage{setspace}
\usepackage{dcolumn}
\usepackage{mathrsfs}
\usepackage{bm}
\usepackage{longtable}

\begin{document}

\preprint{}

\title{Two-dimensional magneto-optical trap as a source for cold strontium atoms}

\author{Ingo Nosske}

\affiliation{Hefei National Laboratory for Physical Sciences at the Microscale and Shanghai Branch, University of Science and Technology of China, Shanghai 201315, China}

\affiliation{CAS Center for Excellence and Synergetic Innovation Center in Quantum Information and Quantum Physics, University of Science and Technology of China, Shanghai 201315, China}

\author{Luc Couturier}

\affiliation{Hefei National Laboratory for Physical Sciences at the Microscale and Shanghai Branch, University of Science and Technology of China, Shanghai 201315, China}

\affiliation{CAS Center for Excellence and Synergetic Innovation Center in Quantum Information and Quantum Physics, University of Science and Technology of China, Shanghai 201315, China}

\author{Fachao Hu}

\affiliation{Hefei National Laboratory for Physical Sciences at the Microscale and Shanghai Branch, University of Science and Technology of China, Shanghai 201315, China}

\affiliation{CAS Center for Excellence and Synergetic Innovation Center in Quantum Information and Quantum Physics, University of Science and Technology of China, Shanghai 201315, China}

\author{Canzhu Tan}

\affiliation{Hefei National Laboratory for Physical Sciences at the Microscale and Shanghai Branch, University of Science and Technology of China, Shanghai 201315, China}

\affiliation{CAS Center for Excellence and Synergetic Innovation Center in Quantum Information and Quantum Physics, University of Science and Technology of China, Shanghai 201315, China}

\author{Chang Qiao}

\affiliation{Hefei National Laboratory for Physical Sciences at the Microscale and Shanghai Branch, University of Science and Technology of China, Shanghai 201315, China}

\affiliation{CAS Center for Excellence and Synergetic Innovation Center in Quantum Information and Quantum Physics, University of Science and Technology of China, Shanghai 201315, China}

\author{Jan Blume}

\affiliation{Hefei National Laboratory for Physical Sciences at the Microscale and Shanghai Branch, University of Science and Technology of China, Shanghai 201315, China}

\affiliation{CAS Center for Excellence and Synergetic Innovation Center in Quantum Information and Quantum Physics, University of Science and Technology of China, Shanghai 201315, China}

\affiliation{Physikalisches Institut, Universit\"at Heidelberg, Im Neuenheimer Feld 226, 69120 Heidelberg, Germany}

\author{Y. H. Jiang}

\email{jiangyh@sari.ac.cn}

\affiliation{Shanghai Advanced Research Institute, Chinese Academy of Sciences, Shanghai 201210, China}

\author{Peng Chen}

\email{peng07@ustc.edu.cn}

\affiliation{Hefei National Laboratory for Physical Sciences at the Microscale and Shanghai Branch, University of Science and Technology of China, Shanghai 201315, China}

\affiliation{CAS Center for Excellence and Synergetic Innovation Center in Quantum Information and Quantum Physics, University of Science and Technology of China, Shanghai 201315, China}

\author{Matthias Weidem\"uller}

\email{weidemueller@uni-heidelberg.de}

\affiliation{Hefei National Laboratory for Physical Sciences at the Microscale and Shanghai Branch, University of Science and Technology of China, Shanghai 201315, China}

\affiliation{CAS Center for Excellence and Synergetic Innovation Center in Quantum Information and Quantum Physics, University of Science and Technology of China, Shanghai 201315, China}

\affiliation{Physikalisches Institut, Universit\"at Heidelberg, Im Neuenheimer Feld 226, 69120 Heidelberg, Germany}

 \date{\today}

\begin{abstract}
We report on the realization of a transversely loaded two-dimensional magneto-optical trap serving as a source for cold strontium atoms. We analyze the dependence of the source's properties on various parameters, in particular the intensity of a pushing beam accelerating the atoms out of the source. An atomic flux exceeding $10^9\,\mathrm{atoms/s}$ at a rather moderate oven temperature of $500\,^\circ\mathrm{C}$ is achieved. The longitudinal velocity of the atomic beam can be tuned over several tens of m/s by adjusting the power of the pushing laser beam.  The beam divergence is around $60$ mrad, determined by the transverse velocity distribution of the cold atoms. The slow atom source is used to load a three-dimensional magneto-optical trap realizing loading rates up to $10^9\,\mathrm{atoms/s}$ without indication of saturation of the loading rate for increasing oven temperature. The compact setup avoids undesired effects found in alternative sources like, e.g., Zeeman slowers, such as vacuum contamination and black-body radiation due to the hot strontium oven.
\end{abstract}


\maketitle

\section{Introduction}
 Two-electron atoms, with the particular example of strontium,  provide unique features for applications in modern atomic physics.  Due to the existence of narrow intercombination lines, they have been successfully used in ultrahigh-precision metrology, e.g., in the realization of the state-of-the-art optical clocks,  with an uncertainty down to the level of $10^{-18}$  \cite{bloom2014,beloy2014}. There has also been increasing interest in ultracold two-electron systems in the fields of quantum gases \cite{takasu2003,kraft2009,deescobar2009,stellmer2009}, quantum simulation \cite{Zhang2014,Zhang2015,Pagano2015,Hofer2015}, quantum information \cite{Daley2008,Gorshkov2009}, ultracold molecules \cite{mcguyer2015,ciamei2017}, and ultracold Rydberg atoms \cite{millen2010,Gil2014,dunning2016}. For the latter application the additional degrees of freedom offered by two active electrons are of importance in Rydberg autoionization \cite{gallagher2005rydberg}, optical dipole trapping of Rydberg atoms \cite{Mukherjee2011} and Rydberg imaging \cite{Lochead2013,McQuillen2013}.

Owing to the advances of laser cooling and trapping, magneto-optical traps (MOTs) have become a standard technique to prepare atomic ensembles at ultralow temperatures. Generally, either Zeeman slowers \cite{Phillips1982,Courtillot2003} or two-dimensional magneto-optical traps (2D-MOTs) (including its variants \cite{Lu1996,Weyers1997,Dieckmann1998}) are commonly used for the initial loading stage of the trap. Compared to a Zeeman slower which requires a dedicated design of the magnetic field configuration and substantial effort in engineering the coil structure, the 2D-MOT is usually preferred due to its more compact size and higher degree of control. A large variety of vapor-cell 2D-MOT designs have been reported in literature. As an example, the vapor-cell variant of a 2D-MOT source is widely used as a loading source for, e.g., rubidium \cite{Dieckmann1998,Schoser2002,Gotz2012}, cesium \cite{Weyers1997,Berthoud2007,Kellogg2012}, potassium \cite{Catani2006} or mercury \cite{petersen2008,witkowski2017dual} MOTs owing to the relatively high vapor pressure of these elements at room temperature.

However, for elements with high melting temperatures, such as the alkaline-earth-metal atoms, a vapor-cell 2D-MOT design can hardly be realized (for a vapor-cell 3D-MOT see \cite{Xu2003}). An alternative is offered by transversely loaded 2D-MOT sources \cite{tiecke2009,lamporesi2013,Dorscher2013}. Unlike the vapor-cell 2D-MOT, which is isotropically loaded from a uniform background vapor, the transversely loaded 2D-MOT captures an atomic beam effusing from a high-temperature oven. Transversely loaded 2D-MOTs are successfully demonstrated for lithium \cite{tiecke2009} and sodium \cite{lamporesi2013} in a compact design, which combines a small distance between the oven and trapping area (typically $10$ cm) with permanent magnets for producing the required 2D quadrupolar magnetic field.  Following up on the seminal work of Tiecke \emph{et al.} \cite{tiecke2009} and Lamporesi \emph{et al.} \cite{lamporesi2013} for alkali-metal atoms, as well as D\"orscher \emph{et al.} \cite{Dorscher2013} for ytterbium, here we present the realization of a compact,  transversely loaded 2D-MOT source for cold strontium atoms.

We characterize the 2D-MOT source by implementing a time-of-flight (TOF) measurement and fluorescence imaging. We measure an atomic flux exceeding $ 10^9\,\mathrm{s}^{-1}$ for $^{88}$Sr at a rather moderate oven temperature of $500\,^\circ\mathrm{C}$. No saturation of the flux for increasing oven temperatures is observed,  which implies that the total atomic flux can be even larger at higher oven temperatures. The divergence of the atomic beam is measured to be around $60\,$mrad, which is larger than the divergence of 2D-MOTs using alkali metals. We find that the divergence of the atomic beam is mainly determined by the larger transverse velocity distribution of Sr atoms in the 2D-MOT, which corresponds to a temperature of $5\,$mK, thus larger than achievable temperatures for alkali-metal atoms. In Section \ref{sec:expsetup} we provide an overview over the apparatus and laser system. Section \ref{sec:flux} is devoted to the characterization of the cold atomic beam emerging from the source. The loading of Sr atoms into a 3D-MOT is described in Section \ref{sec:loading}. Section \ref{sec:conclusion} contains concluding remarks on future improvements and a comparison of our source to alternative approaches.

\section{2D-MOT setup}\label{sec:expsetup}

\begin{figure*}[t]
  \centering
  \includegraphics[width=0.9\linewidth]{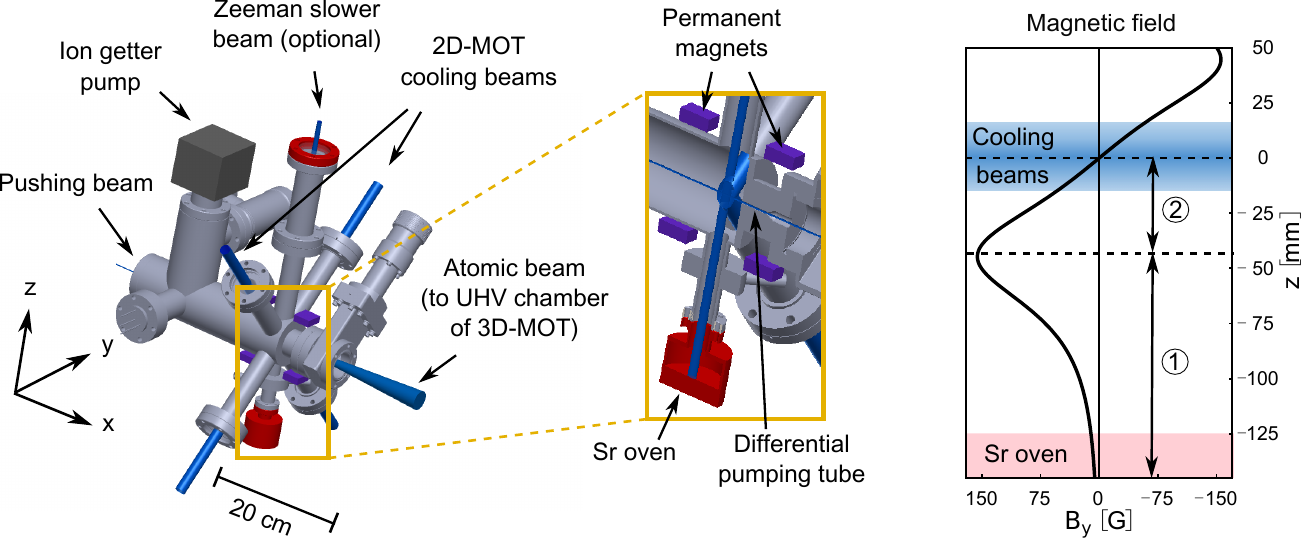}
  \caption{(color online) Schematic of the experimental setup for the transversely loaded 2D-MOT. Left: Experimental setup. Hot strontium atoms effusing from the oven in the bottom of the setup are cooled and trapped in the center of the multiway cross vacuum chamber by a perpendicular pair of retroreflected cooling beams. Atoms in the 2D-MOT are pushed to a UHV chamber (not shown) where they are further loaded into a 3D-MOT. A Zeeman slower beam (optional) is used to decelerate hot strontium atoms through the top viewport of the setup.  Heated parts of the setup are highlighted in red. The vacuum of the setup is maintained by an ion getter pump. Center: Zoom-in of the highlighted rectangle in the left part. Four stacks of permanent magnets are used to generate the 2D quadrupolar magnetic field. A differential pumping tube is located between the 2D-MOT vacuum chamber and the UHV chamber for vacuum isolation. Right: Magnetic field along the $z$ direction. The red shaded area depicts the strontium oven, while the blue shaded area represents the 2D cooling beams. For Zeeman deceleration, either region \textcircled{\small{1}} or \textcircled{\small{2}} can be used.}
\label{fig:vacsys}
\end{figure*}

\subsection{Vacuum chamber}

The schematic of our experimental setup is shown in Fig.~\ref{fig:vacsys}. A strontium oven is located at the bottom ($z$ direction) of a custom-made multiway cross vacuum chamber (LEOSolutions). Hot strontium atoms evaporated from the oven are cooled and trapped in the center of the multi-way cross vacuum chamber by a perpendicular pair of retroreflected laser beams. A weak laser beam (pushing beam) propagating along the $x$ direction pushes atoms in the 2D-MOT to a small orifice separating the source chamber from a subsequent UHV chamber hosting the 3D-MOT (not shown). The 2D quadrupole magnetic field is generated by four stacks of permanent magnets, which also creates a magnetic-field gradient along the propagation direction of the hot atomic beam. Such a magnetic field design enables one to implement an auxiliary Zeeman slower by applying an additional laser beam through a viewport opposite to the strontium oven, as first demonstrated in Ref.~\cite{lamporesi2013}. A differential pumping section is inserted to isolate the vacuum chamber from the UHV chamber, which has an inner diameter of $2\,$mm with a length of $22.8\,$mm. Its entry has no direct line of sight to the oven, and no clogging of the tube is noticed by observation through the pushing beam viewport.

The entire apparatus follows a very compact design. The distance from the strontium oven aperture to the 2D-MOT trapping region is only $125$ mm [see Fig.~\ref{fig:vacsys} (center)], and is thus significantly smaller than the typical length of strontium Zeeman slowers ($\sim300$ mm) \cite{Courtillot2003,Bober2010}. The absence of coil to generate magnetic fields greatly simplifies the setup. A small yet efficient ion getter pump (SAES Getters, NEXTorr D 200-5) is mounted to maintain the system vacuum pressure below $10^{-10}\,$mbar when the Sr oven is not heated ($\sim 2\times10^{-10}\,$mbar at $465\,^\circ\mathrm{C}$). We note that a very compact integrated solution of strontium Zeeman slower source is offered by AOSense (see Supplementary Materials of Ref.~\cite{norcia2016}).

In most strontium Zeeman slowers, the oven temperature has to be ramped up to $\sim 550-600\,^\circ\mathrm{C}$ to achieve a sufficient atomic flux because of the rather high melting point of strontium. In addition, to reduce the atomic beam divergence, an array of hundreds of microtubes is frequently mounted in the oven aperture to enhance the atomic beam collimation. However, in the case of our 2D-MOT approach, the strontium oven is operated in the temperature range between $450$ and $510\,^\circ\mathrm{C}$ without a microtube array, due to the closer distance to the trapping area. The oven is a simple cylinder with an aperture of $16\,$mm diameter, where 5\,g of strontium with natural abundance ($99.99\%$ purity; Sigma \& Aldrich) are deposited. It is heated to $465~^\circ\mathrm{C}$ for the measurements presented in this paper, if not stated otherwise. To avoid coating of the viewport for the auxiliary Zeeman slower by the flux of hot strontium atoms, the window, made of sapphire crystal, is permanently heated to $330~^\circ\mathrm{C}$. Yet, we still observe slight degradation of the laser transmission through this window in the course of several months of operation.

To create the required 2D axial symmetric quadrupolar magnetic field, N35 neodymium ($\mathrm{Nd}_2\mathrm{Fe}_{14}\mathrm{B}$) magnets are used. Each single piece of the magnet has dimensions of 25~mm $\times$ 10~mm $\times$ 3~mm and a magnetization of $6.6(1) \times 10^5~\mathrm{A/m}$ along the shortest dimension. As shown in Fig.~\ref{fig:vacsys}, four stacks of permanent magnets are located symmetrically around the chamber, with a distance of $75\,$mm and $88\,$mm in $x$ and $z$ directions, respectively. Each stack consists of nine magnets glued together. The magnetization axes of the two upper stacks (positive $y$ direction) point against the two lower stacks (negative $y$ direction). The resulting field is sketched in Fig. 2 of Ref.~\cite{lamporesi2013}. With such a configuration, the permanent magnets can generate a quadrupole magnetic field for 2D-MOT cooling beams up to $50\,$G/cm (for the specific configuration given above, the gradient amounts to $34\,$G/cm in the trapping area).

In addition, they also produce a gradient magnetic field ($\partial B_y/ \partial z $) in the $z$ direction for an auxiliary Zeeman slower. However, because the magnetic field is pointing towards the $y$ direction, only half the power of the Zeeman slower beam can actually be used for deceleration since its linear polarization is oriented along the $x$ direction. There are two regions of the magnetic field topology allowing for the Zeeman slower, as labeled \textcircled{\small{1}} and \textcircled{\small{2}} in Fig.~\ref{fig:vacsys} (right). The magnetic field gradient is around $15\,$G/cm in the region \textcircled{\small{1}} and it is twice as large in the region \textcircled{\small{2}}.

\subsection{Laser system}

\begin{figure}
\centering
\includegraphics[width=0.9\linewidth]{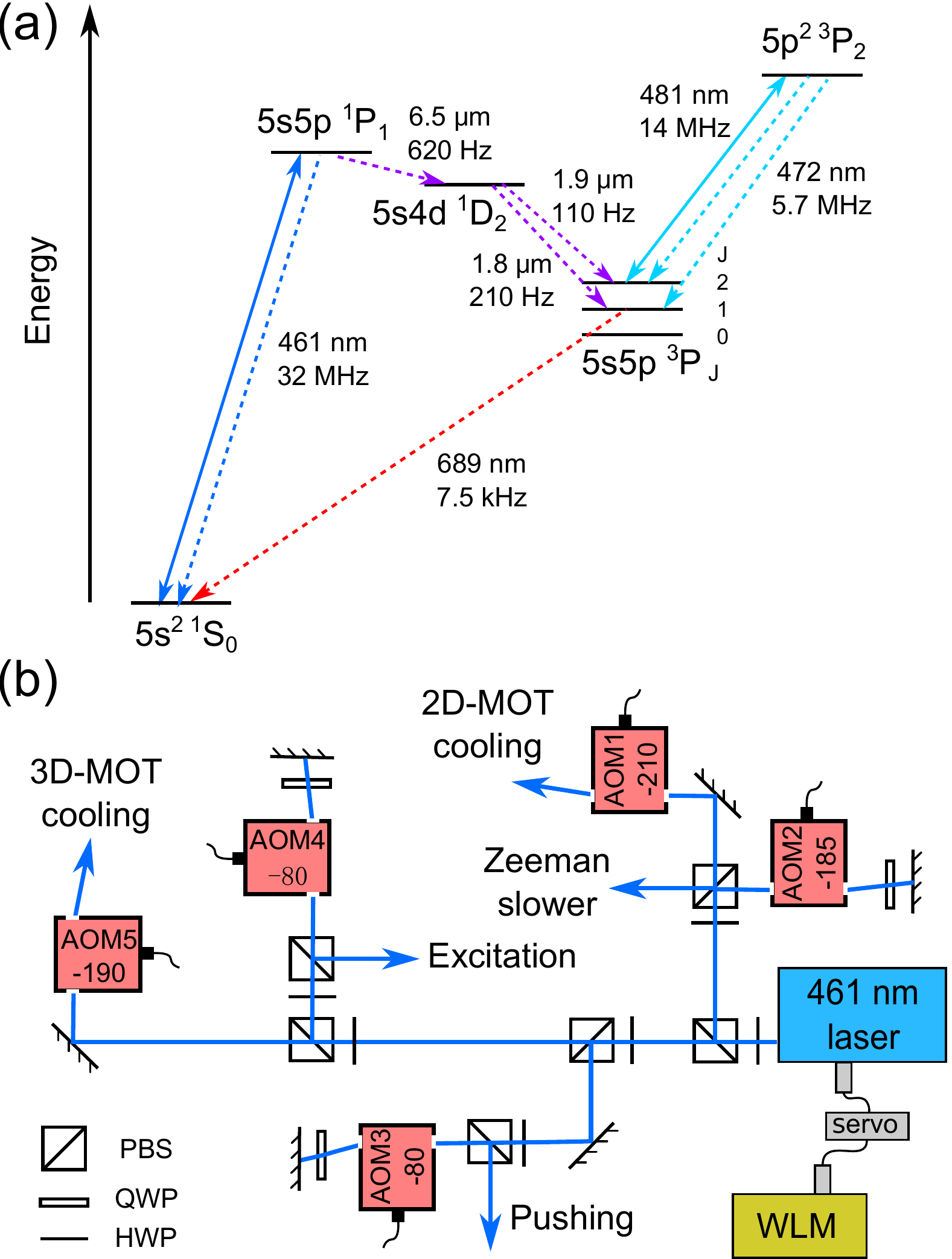}
 \caption{(color online)  Energy-level scheme and schematic of the laser setup for the strontium 2D-MOT.  (a) Energy diagram of the bosonic strontium. $5s5p\,^3\mathrm{P}_2$ is a dark state in the laser cooling and trapping. It is also a reservoir state for the magnetic-field trap. Spectroscopic parameters are taken from Ref.~\cite{sansonetti2010}; decay rates involving $5s4d\,^1\mathrm{D}_2$ can be found in Ref.~\cite{Xu2003}. (b) Schematic of the laser setup. A frequency-doubled Ti:sapphire laser system at a wavelength of 461 nm is used to provide all laser beams for the experiment. The Ti:sapphire laser frequency is stabilized by a High Finesse wavemeter with a servo loop. QWP: quarter-wave plate; HWP: half-wave plate; WLM: wavemeter.}
\label{fig:energylevels}
\end{figure}

Figure~\ref{fig:energylevels}(a) shows the relevant energy levels of bosonic strontium isotopes for laser cooling and trapping. Unlike alkali-metal atoms, bosonic isotopes of alkaline-earth-metal atoms have no nuclear spin so that there is no hyperfine structure.  Therefore, the singlet transition of $ 5s^2 \,^1\mathrm{S}_0 \rightarrow 5s5p\,^1\mathrm{P}_1$ ($461\,$nm) represents an almost perfect realization of a two-level system serving as an efficient cycling transition for laser cooling, with only a weak spontaneous decay channel to the $5s4d\,^1\mathrm{D}_2$ state. As the spontaneous decay rate of $5s5p\,^1\mathrm{P}_1$ is $\Gamma/2\pi = 32\,$MHz, the associated Doppler temperature ($770\,$$\mu$K) is higher than that of, e.g., rubidium ($146\,$$\mu$K).

Due to the small loss rate ($\sim 100\,$Hz) from the decay channel, i.e., $ 5s5p\,^1\mathrm{P}_1 \rightarrow 5s4d\,^1\mathrm{D}_2 \rightarrow 5s5p\,^3\mathrm{P}_2 $, where $5s5p\,^3\mathrm{P}_2 $ is a reservoir state with a lifetime on the order of $100\,$s, laser beams to repump population into the cooling cycle are generally not necessary for the strontium 2D-MOT. However, applying repumping light can play an important role in prolonging the lifetime of a 3D-MOT (see Sec. \ref{sec:loading}) or by recollecting atoms in the reservoir state magnetically trapped by the MOT field \cite{Xu2003,stellmer2014b}. Accordingly, we use a repumping laser (Toptica, DL 100) to address the transition $5s5p\,^3\mathrm{P}_2 \rightarrow 5p^2\,^3\mathrm{P}_2 $ at $481\,$nm to increase the lifetime of the 3D-MOT \cite{stellmer2014a,stellmer2014b}. An order of magnitude increase of the atom number is observed in our experiment.

We use a Ti:sapphire laser and a frequency doubler (MSquared, SolsTiS and ECD-X) to generate the blue laser light for the cooling transition at $461\,$nm. The laser system typically delivers a power of $500\,$mW after the frequency doubler. For frequency stabilization, the Ti-sapphire laser light at $922\,$nm is coupled to a high-accuracy wavemeter (High Finesse, WSU/10) via a single-mode fiber. The output signal of the wavemeter serves as input to a servo loop controlling the frequency of the laser. When locked, the remaining frequency fluctuations of the blue laser light are measured to be $1.1 \,$MHz (rms) averaged over a time interval of $11\,$h, thus much smaller than the natural linewidth of the atomic cooling transition. Details on the stabilization scheme will be published elsewhere.

The blue laser output is coupled to an optical path system consisting of acousto-optical modulators (AOMs), polarizing beam splitters (PBSs), waveplates, and fiber couplers to deliver all the laser frequencies and powers involved in the experiment, such as the 2D-MOT cooling beam, the Zeeman slower beam, the pushing beam, etc. [see Fig.~\ref{fig:energylevels}(b)]. In the experiments described below, the 2D-MOT cooling beam, the Zeeman slower beam and the pushing beam have $1/e^2$ radii of $7.5\,$mm, $6\,$mm, and $0.7\,$mm, respectively. Their detuning from the atomic resonance is typically $-1.5~\Gamma$, $-7~\Gamma$, and zero, respectively, and their peak intensities are around $4~I_\mathrm{sat}$ (four beams combined), $I_\mathrm{sat}$, and $0.2~I_\mathrm{sat}$, respectively. Here $I_\mathrm{sat}$ denotes the saturation intensity of the transition and is given by 43\,mW/cm$^2$. The Zeeman slower beam is slightly focused at the position of the 2D-MOT trapping area, and its detuning corresponds to the decreasing field configuration. For the 2D-MOT cooling beam alignment, we use the $\sigma^+-\sigma^-$ configuration, constructed by two perpendicular pairs of retro-reflected laser beams with circular polarizations.

\section{Cold atomic beam properties}\label{sec:flux}

\begin{figure}[h]
\centering
	\includegraphics[width=0.9\linewidth]{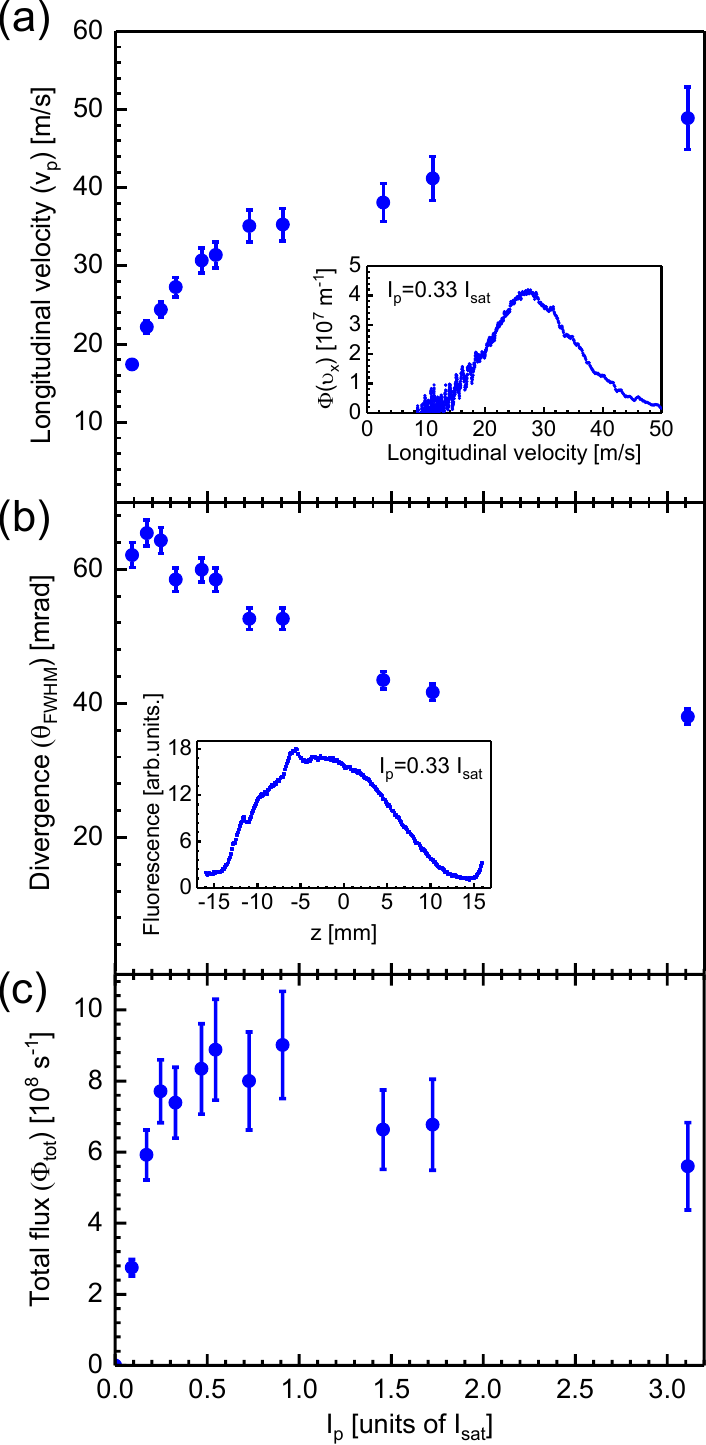}
  \caption{ (color online) The most probable longitudinal velocity, divergence, and the total flux of the atomic beam versus the pushing beam light intensity, respectively. The insets in (a) and (b) show the longitudinal velocity distribution and the transverse distribution of the atomic beam at $I_\mathrm{p} = 0.33\,I_\mathrm{sat}$. The field gradient is chosen to $50~\mathrm{G/cm}$, the oven temperature is $465~^\circ\mathrm{C}$, and Zeeman slower loading was applied (see Sec. \ref{sec:Zeeman}).}
	\label{fig:coldbeam}
\end{figure}

We first characterize the cold atomic flux of the strontium 2D-MOT source by the TOF measurement \cite{Dieckmann1998}. We apply a retroreflected laser beam on resonance with the singlet transition ($^{88}$Sr) to excite the cold atomic beam emerging from the 2D-MOT. The linearly polarized excitation beam (not shown in Fig.~\ref{fig:vacsys}) propagates along the $z$ direction at a distance of $300\,$mm downstream from the center of the 2D-MOT. It has a $1/e^2$ radius of $2.4\,$mm at a power of $6\,$mW, corresponding to a maximum saturation parameter of two beams $s = I/I_\mathrm{sat} = 3$. The atomic fluorescence during the excitation is collected with a photodiode and additionally monitored by a complementary metal-oxide semiconductor (CMOS) camera. When rotating the direction of the linear polarization of the excitation beam, the fluorescence signal is found to simply follow the classical dipole radiation pattern, as expected for this simple effective two-level system.

The temporal evolution of the fluorescence signal is measured after suddenly turning off the pushing laser beam, from which we then determine the information about the atomic flux. The atomic flux distribution on longitudinal velocity is then given by

\begin{eqnarray}\label{eq:fluxformula}
\Phi(v_x) = -\eta \frac{\tilde{A}\ell}{v_x}\frac{dU(t)}{dt},
\end{eqnarray}

where $v_x$ is the longitudinal velocity, $\ell$ is the distance from the 2D-MOT to the excitation beam, $U(t)$ is the voltage signal measured from the fluorescence, and $\eta$ is a calibration factor taking the overall detection efficiency, excitation efficiency, and transimpedance gain of the fluorescence detector into account (see the Appendix for details). $\tilde{A}$ is the effective cross section of the cold atomic beam at the excitation area, which can be deduced from the measurement of beam divergence. By velocity integration of Eq.~(\ref{eq:fluxformula}), the total atomic flux $\Phi_\mathrm{tot}$ could be determined.

We measure different parameters of the cold atomic beam versus the pushing beam light intensity in Fig.~\ref{fig:coldbeam}. $I_p$ denotes the peak intensity, as also in the following figures. The most probable velocity $v_p$ is obtained from the measurement of $\Phi(v_x)$ at each value of $I_p$ [see the inset of Fig.~\ref{fig:coldbeam}(a)]. It varies from $17\,$m/s to $49\,$m/s with increasing $I_p$, and appears to increase at a smaller rate beyond $I_\mathrm{sat}$.

The divergence angle $\theta_{\mathrm{FWHM}}$ corresponding to the full width at half maximum of the transverse distribution of the atomic beam is analyzed by fluorescence imaging with the CMOS camera. As an example, the inset of Fig.~\ref{fig:coldbeam}(b) shows a typical measurement of transverse distribution of the atomic beam. As seen in Fig.~\ref{fig:coldbeam}(b), $\theta_{\mathrm{FWHM}}$ decreases with increasing $I_p$, which can be understood by a simple model of the atomic cloud expansion based on two effects which govern the expansion process. The first one is the initial transverse velocity distribution of the 2D-MOT beam, which dominates the expansion process at larger $I_p$. The Doppler temperature of $1.4\,$mK [taking laser detuning and the value of $v_p$ in Fig.~\ref{fig:coldbeam}(a) inset into account] yields a beam divergence of $26\,$mrad.  As $v_p$ increases with increasing of $I_p$ [see Fig.~~\ref{fig:coldbeam}(a)], the expansion time of the atomic beam becomes less to reach the detection area for larger $I_p$. Therefore, the divergence of atomic beam shows a decreasing trend with increasing $I_p$. From the dependence of $\theta_{\mathrm{FWHM}}$ at larger $I_p$, we estimate that the transverse velocity distribution corresponds to a temperature of $5\,$mK, which is larger than the Doppler temperature of $1.4\,$mK.  As an additional factor, the geometrical size of the differential pumping tube ($14\,$mm away from the center of the 2D-MOT) limits the beam divergence to roughly $60\,$mrad, which explains the measured beam divergence at small $I_p$.

Through the above measurements of longitudinal velocity distribution and beam divergence, we deduce the total atomic flux $\Phi_\mathrm{tot}$ as a function of pushing beam intensity [see Fig.~\ref{fig:coldbeam}(c)] using Eq.~(\ref{eq:totalflux}) in the Appendix. $\Phi_\mathrm{tot}$ rises to $9\times 10^8\,\mathrm{atoms/s}$ with increasing $I_p$. Beyond saturation intensity the flux decreases to about $6\times 10^8\,\mathrm{atoms/s}$. The rising of $\Phi_\mathrm{tot}$ is correlated with the increase of $v_p$ [see Fig.~\ref{fig:coldbeam}(a)], resulting in an increase of the total atomic flux due to Eq.~(\ref{eq:totalflux}) in the Appendix. However, although $v_p$ still increases with a larger $I_p$, the total atomic flux no longer continues to rise but instead decreases slightly. The reason is that the cooling process of the 2D-MOT is perturbed by the force exerted by the pushing beam. The optimum flux is achieved around the saturation intensity of the pushing beam light intensity.

\section{Loading of a strontium 3D-MOT}\label{sec:loading}

\begin{figure}
\centering
	\includegraphics[width=0.9\columnwidth]{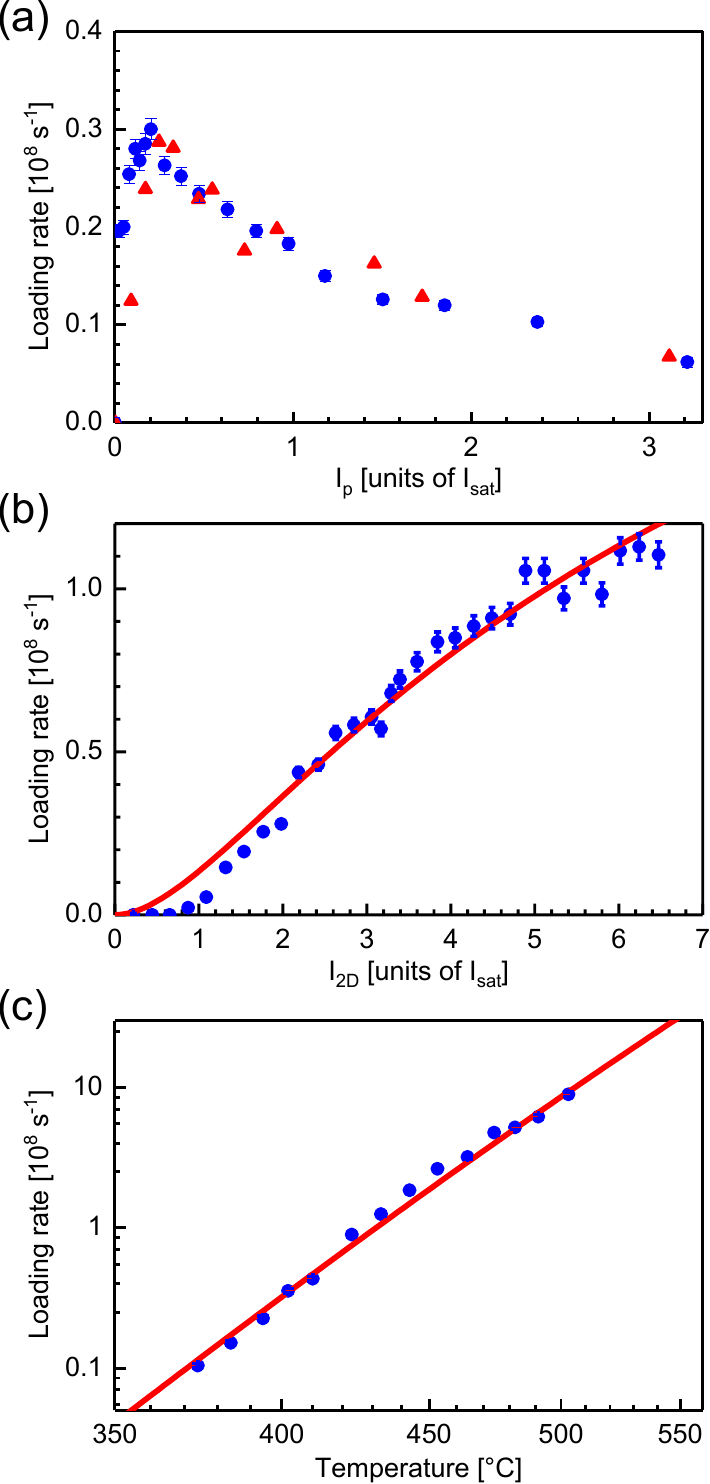}
  \caption{(color online) 3D-MOT loading rate as a function of the pushing beam light intensity, the 2D-MOT cooling beam light intensity, and the oven temperature. For each panel, the unvaried parameters are set to $0.2~I_\mathrm{sat}$, $4~I_\mathrm{sat}$ (four beams combined), and $465~^\circ\mathrm{C}$, respectively. (a) Loading rate versus the pushing beam light intensity. The triangles indicate results of a model calculation. (b) Loading rate versus the 2D-MOT cooling beam light intensity.  The solid line represents a fit with a model. (c) Loading rate versus the oven temperature (using flux enhancement of the 2D-MOT by the decreasing field Zeeman slower for this set of measurements). The solid line represents a model calculating the flux from the strontium vapor pressure and the 2D-MOT capture efficiency. Details on all model calculations can be found in the text.}
	\label{fig:loadrate}
	\end{figure}

The cold beam of atoms is used to load a 3D-MOT located in a UHV chamber 335\,mm downstream. The setup, procedures, and parameters of the 3D-MOT closely follow methods established by other groups (see, e.g., Refs.~\cite{Xu2003,stellmer2009}). In our setup, the magnetic-field gradient of the 3D-MOT along $z$ is typically set to 40~G/cm, the combined trapping beam intensity is 2.5 $I_\mathrm{sat}$ (six beams combined, $1/e^2$ radius 6\,mm) and the detuning is chosen around -1 $\Gamma$.

Figure~\ref{fig:loadrate}(a) plots the loading rate as a function of the pushing beam light intensity. The loading rate reaches a peak at pushing beam intensities around $I_p=0.2\,I_\mathrm{sat}$, beyond which it decreases rapidly as $I_p$ becomes larger. The red triangles depict a model calculation, in which we integrate the atomic flux as measured in Fig.~\ref{fig:coldbeam} assuming a finite capture velocity and acceptance solid angle of the 3D-MOT [see Eq.~(\ref{eq:totalflux}) in the Appendix]. The latter is determined by the size of the 3D-MOT's laser beams. Thus, the only free parameter in the model is the capture velocity of the 3D-MOT. As the loading flux of the 3D-MOT was measured under slightly different conditions as the data given in Fig.~\ref{fig:coldbeam} concerning 2D-MOT parameters, we have also included a scaling parameter of the total loading rate which we find to range between 0.2 and 2. Comparison with measured data as shown, e.g., in Fig.~\ref{fig:loadrate}(a), yields capture velocities between $20\,\mathrm{m/s}$ and $30\,\mathrm{m/s}$, which lies slightly below an estimation based on the actual parameters of the 3D-MOT laser intensities and beam sizes.

From this model, we deduce that the initial increase of the loading flux reflects the increasing total flux [see Fig.~\ref{fig:coldbeam}(c)], while the decrease at larger pushing beam intensities results from the finite capture efficiency of longitudinal velocities beyond the capture velocity [see Fig.~\ref{fig:coldbeam}(a)]. The estimated capture efficiency at maximum loading rate is around $10\%$. By increasing the power and diameter of the 3D-MOT cooling beams,  the capture efficiency could be further enhanced, in principle. However, constraints of our setup, in particular concerning the laser power available and geometric constraints of the laser beam sizes, currently impede significant improvements.

We also measure the loading rate as a function of the 2D-MOT cooling beam light intensity, as depicted in Fig.~\ref{fig:loadrate}(b). The loading rate rises monotonically with increasing 2D-MOT cooling beam light intensity beyond a threshold value of about $~I_\mathrm{sat}$. At larger intensities, the loading rate begins to saturate. The red curve in Fig.~\ref{fig:loadrate}(b) is fitted by a simple model considering the capture velocity dependence on the cooling beam light intensity of the 2D-MOT.  The total atomic flux of the 2D-MOT source scales with the fourth power of the capture velocity ($\Phi\sim v^4_c$), which is a function of the cooling beam light intensity $I_\mathrm{2D}$ as $ v^2_c \sim s/(s+1)$, $s=I_\mathrm{2D}/I_\mathrm{sat}$ \cite{monroe1990,tiecke2009}. While this model quantitatively captures the essential trend of the observed data, it fails to qualitatively reproduce the observed threshold of the loading rate at smaller intensities.

The loading rate versus the oven temperature is shown as a double-logarithmic plot in Fig.~\ref{fig:loadrate}(c). The power-law behavior can be reproduced by simply assuming that the loading rate follows the vapor pressure of Sr atoms in the oven. The solid curve in Fig.~\ref{fig:loadrate} is given by $\Phi\sim P_\mathrm{Sr}(T)\times T^{-5/2}$, where $P_\mathrm{Sr}(T)$ is the saturated vapor pressure of solid strontium at a temperature $T$ \cite{alcock1984}. The additional factor of $T^{-5/2}$ stems from the temperature dependence of the captured atomic flux and  total atomic flux, as can be understood as follows. The normalized velocity distribution of the hot atomic beam follows a Maxwell-Boltzmann distribution of particle flux \cite{Beijerinck1975}. For a small capture velocity $v_c$,  the fraction of velocity distribution of the hot atomic flux within the range of capture velocity scales with the mean velocity $\overline{v}$ as $(\overline{v})^{-4}$. Therefore, with the increase of the oven temperature, the fraction of hot atoms captured in the 2D-MOT scales with $T^{-2}$. For a given vapor pressure,  the total atomic flux of the hot atoms is proportional to the product of the mean velocity ($T^{1/2}$) and atomic density ($T^{-1}$), which gives a scaling of $T^{-1/2}$, finally resulting in a $T^{-5/2}$ dependence. Comparing to this model, we conclude that there is no additional loss mechanism up to oven temperatures of  $500\,^\circ\mathrm{C}$, where we obtain the maximum loading rate of $9\times10^8\,\mathrm{s^{-1}}$ in our experiment. We expect to achieve much higher loading flux by further increasing oven temperature, but so far did not dare to do so in order to avoid rapid depletion of the oven or possible damage to our apparatus.

\section{Effect of Zeeman slowing}\label{sec:Zeeman}

Motivated by the results for a sodium 2D-MOT reported in Ref.~\cite{lamporesi2013}, we investigated the effect of an auxiliary Zeeman slower realized in the fringe fields of the magnet configuration (see the right graph in Fig.~\ref{fig:vacsys}). Compared to an optimal gradient of $\sim20\,$G/cm and a much longer working distance of $300\,$mm of strontium Zeeman slowers in the literature \cite{Courtillot2003,Bober2010}, the region \textcircled{\small{1}} should be better suited for slowing strontium atoms. Nevertheless, we find that region \textcircled{\small{2}} is more effective for Zeeman slowing. One possible reason is that the slowing effect is disrupted in region \textcircled{\small{1}} before atoms are trapped in the 2D-MOT, due to collisions with surrounding hot atoms.

By switching the Zeeman slower beam on and off, we measure the enhancement factor of the loading rate into the 3D-MOT with and without the effect of Zeeman slowing. Using the decreasing slope of the magnetic field (indicated as region \textcircled{\small{2}} in Fig.~\ref{fig:vacsys}), a maximum enhancement of $4$ is observed at an optimum detuning of $\Delta_{Z}=-210\,$MHz (corresponding to $-6.6\,\Gamma$) and a maximum available laser intensity of $I_{Z}=1.2\,I_\mathrm{sat}$ ($1/e^2$ radius $6\,$mm). We find that for a smaller detuning, the enhancement factor is generally smaller and saturates at a lower light intensity. For example, at the detuning of $\Delta_{Z}=-175\,$MHz ($-5.5\,\Gamma$), we observe an enhancement factor of $3$ at a low intensity of $I_{Z}=0.3\,I_\mathrm{sat}$, where it already shows clear signs of saturation versus power.  At the increasing slope of the magnetic field (region \textcircled{\small{1}} in Fig.~\ref{fig:vacsys}) we also find an effect of Zeeman slowing, yet with a smaller enhancement (factor of 2). As a comparison, using the increasing slope of the magnetic field Lamporesi \emph{et al.}~\cite{lamporesi2013} measured an enhancement factor of $12$ at laser intensities corresponding to an order of magnitude higher saturation for the Zeeman slower and larger beam diameters capturing more atoms. Therefore, we would expect a larger enhancement effect of the Zeeman slower if more laser power was available in our setup.

\section{Conclusion}\label{sec:conclusion}

We have realized a transversely loaded 2D-MOT as a source for cold strontium atoms, which is compact and easy to implement into a more complex experimental setting.  The 2D-MOT source can generate an atomic flux exceeding $10^9\,\mathrm{atoms/s}$ (for the isotope $^{88}$Sr) at a rather moderate oven temperature of $500\,^\circ\mathrm{C}$. Both the longitudinal velocity distribution and divergence of the cold atomic beam are found to limit the loading efficiency of the 3D-MOT, which is around $10\%$ in our current setup. One possible way to increase the capture efficiency is to enlarge the waist of the 3D-MOT laser beams at the expense of assigning more laser power to the 3D-MOT. Also, a steady increase of the loading rate with the light intensity of the 2D-MOT is observed showing only weak indications of saturation.

The decreasing slope of the gradient magnetic field has been observed to be efficient for realizing an auxiliary Zeeman slower to enhance the capture of atoms into the 2D-MOT. As the maximum of the magnetic fringe is reached at around 50 mm from the 2D-MOT location (see Fig. ˜\ref{fig:vacsys}), one might consider placing the strontium oven at an even shorter distance than currently realized. Thus, the atomic flux into the 2D-MOT could be further increased. This feature might be advantageous for the design of miniaturized cold strontium sources for applications, e.g., in satellite-based atomic clocks or atom interferometers.

\begin{table*}[t]
		\centering
			\begin{tabular}{p{1.8cm}p{1.8cm}p{2.7cm}p{2.9cm}p{1.8cm}p{1.8cm}p{1.8cm}}
		  \hline \hline Isotope & Statistics & Abundance \cite{sansonetti2010} & $\Delta f$ (MHz) \cite{eliel1983,kluge1974} & $L\, (10^6\,\mathrm{atoms/s}$) &\quad\, $L_\mathrm{rel}$ \\ \hline
			$^{88}$Sr & Bosonic & $82.58\%$ & 0  & 300(11)  &  \quad\,1\\
			$^{87}$Sr & Fermionic & $7.00\%$ & -51.9  & 22(1) & \quad\,0.87(6)\\
			$^{86}$Sr & Bosonic & $9.86\%$ & -124.8  & 38(1) & \quad\,1.06(5)\\
			$^{84}$Sr & Bosonic & $0.56\%$ & -270.8  & 2.2(5)  & \quad\,1.0(3)\\
			\hline\hline
			\end{tabular}\\
		\caption{Overview of loading rates of all the stable isotopes of strontium (using flux enhancement of the 2D-MOT by the decreasing field Zeeman slower for this set of measurements). $\Delta f$ is the relative frequency shift to $^{88}$Sr for the singlet cooling transition. $L_\mathrm{rel}$ shows the relative loading rate scaled to the natural abundance. }
		\label{tab:fluxoverview}
		\end{table*}

By adjusting the laser frequencies appropriately, we have operated the 2D- and 3D-MOTs under conditions to capture each of the stable isotopes of Sr. The loading rates of all the stable isotopes of strontium at our typically used oven temperature of $465~^\circ\mathrm{C}$ are summarized in Table~\ref{tab:fluxoverview}. We find that for the bosonic isotopes the loading rates in this experimental run are consistent with the natural abundancies, which indicates that the physical processes of the 2D-MOT and 3D-MOT do not depend on the isotope, as expected from the simple level structure. The slightly lower loading rate for fermionic $^{87}$Sr is likely due to its unresolved hyperfine structure on the cooling transition.

Currently, the effect of blackbody radiation emerging from an oven is one of the major limiting factors to the performance of optical clocks based on strontium atoms \cite{Safronova2012}. To avoid such problems in current setups using Zeeman slowers, a mechanical shutter can be installed to temporally block the atomic beam or use beam deflector \cite{norcia2016,Yang2015}. As an advantage, the transversely loaded 2D-MOT source neither causes contamination in the ultrahigh vacuum chamber nor produces significant blackbody radiation, because there is no direct line of sight from the high-temperature oven to the area in the UHV chamber, where the experiments with the trapped ultracold atoms are performed.

The atom flux could be further enhanced by increasing the oven temperature, as we do not observe any saturation of the loading flux with the increase of oven temperature. For example, the atomic flux for $^{88}$Sr would increase to the $10^{11}\,\mathrm{atoms/s}$ range at temperatures around $600\,^\circ\mathrm{C}$, as estimated from the model described above. The corresponding loading rates would then be comparable to those of most reported strontium 3D-MOTs using Zeeman slowers~\cite{Hill2014,Yang2015,Bennetts2017}.

\textit{Note added in Proof.} We recently became aware of a 2D-MOT based source of cold strontium atoms, which is employed to load a trapped-ion array \cite{bruzewicz2016}.

\section*{Acknowledgments}

We thank G. Z\"urn, J. Ulmanis, D. Litsch, and A. Salzinger for help in planning the experiment and useful discussions. We are grateful to F. Schreck for recommending the repumper transition and other helpful remarks.  M.W.'s research activities in China are supported by the 1000-Talent-Program of the Chinese Government. The work was supported by the National Natural Science Foundation of China (Grants No.~11574290 and No.~11604324). Y.H.J. also acknowledges NSFC support under Grants No.~11420101003 and No.~91636105.

\section*{Appendix}
The measured fluorescence voltage signal $U(t)$ in the TOF measurement is given by
\begin{equation}\label{eq:fluxformula2}
U(t)=\hat{\eta}\frac{\hbar\omega\Gamma}{2 }S_d V n(t),
\end{equation}
where $\hbar$ is the reduced Planck constant, $\omega$ is the angular frequency of the fluorescence light,  $\Gamma$ is the spontaneous decay rate of the $5s5p\,^1\mathrm{P}_1$ state, $S_d$ is the effective saturation parameter of the excitation beam within the detection volume $V$, and $n(t)$ is the atomic density. $\hat{\eta}=R\,G\,\Omega_\mathrm{PD}\,T_\mathrm{opt}F_\mathrm{dip}$ is a product of the photodiode sensitivity $R$, the current-to-voltage transimpedance gain $G$, the solid angle of detection $ \Omega_\mathrm{PD}$, the optical loss $T_\mathrm{opt}$, and the weighting factor $F_\mathrm{dip}$ due to the dipole radiation pattern.

The time derivative of $n(t)$ with time can be expressed as
\begin{equation}\label{eq:fluxformula3}
\frac{dn(t)}{dt}=-\frac{v_x^2}{\ell}n_v(v_x),
\end{equation}

where $n_v(v_x)$ is the atomic density distribution on the longitudinal velocity $v_x$, with $v_x = \ell/t$.

Substituting Eq.~(\ref{eq:fluxformula3}) into Eq.~(\ref{eq:fluxformula2}), the time derivative of $U(t)$ can be expressed as
\begin{equation}
-\frac{dU(t)}{dt}=-\frac{\hat{\eta}\hbar\omega\Gamma}{2 }S_d V \frac{dn(t)}{dt}=v_x^2\frac{\hat{\eta}\hbar\omega\Gamma }{2 \ell }S_d V n_v(v_x).
\end{equation}
Therefore, the longitudinal velocity distribution of the atomic beam within a divergence angle $\theta$ can be written as
\begin{equation}\label{eq:fluxformula4}
\Phi(v_x) =\tilde{A}v_x n_v(v_x) = -2\frac{\tilde{A} \ell }{v_x \hat{\eta}\hbar\omega\Gamma S_d V}\frac{dU(t)}{dt},
\end{equation}

where $\tilde{A}=A \int_0^{\Omega_c} g{(\theta)} d\Omega$ is the effective cross section of the atomic beam at the detection position.  $A$ is the cross section of the total atomic flux. $\Omega_c$ is the acceptance solid angle of the atomic beam in the detection area. $g(\theta)$ is the normalized transverse distribution of the atomic beam. From Eq.~(\ref{eq:fluxformula4}), the calibration factor $\hat{\eta}=2/(\eta\hbar\omega\Gamma S_d V)$.

Then the total atomic flux is given by the integration

\begin{equation}\label{eq:totalflux}
\Phi_\mathrm{tot} =\int A \, v_x\, n_v(v_x) dv_x.
\end{equation}

\bibliography{2D_MOT_references}

\end{document}